\documentclass[preprint,...]{revtex4-1}

\usepackage{graphicx}

\usepackage{amsmath}
\usepackage{amsfonts}
\usepackage{amssymb}
\usepackage{calrsfs}

\usepackage{appendix}
\usepackage{emptypage}
\usepackage{cancel}
\usepackage{fancyvrb}
\usepackage{gensymb}
\usepackage{verbatim}
\usepackage{wrapfig}
\usepackage{units}
\usepackage{bm}
\usepackage{url}
\usepackage{xfrac}
\usepackage{xcolor}
\usepackage[english]{babel}
\usepackage{blindtext}
\usepackage{setspace}
\usepackage{csquotes}

\usepackage{array}
\newcolumntype{L}[1]{>{\raggedright\let\newline\\\arraybackslash\hspace{0pt}}m{#1}}
\newcolumntype{C}[1]{>{\centering\let\newline\\\arraybackslash\hspace{0pt}}m{#1}}
\newcolumntype{R}[1]{>{\raggedleft\let\newline\\\arraybackslash\hspace{0pt}}m{#1}}

\newcommand{\abs}[1]{\left| #1 \right|} 
\newcommand{\pd}[2]{\frac{\partial #1}{\partial #2}} 
 
\let\baraccent=\= 
\renewcommand{\=}[1]{\stackrel{#1}{=}} 

\begin{document}
	
	
	\title{Intrinsic suppression of turbulence in linear plasma devices}
	
	
	\author{J.Leddy$^{1,2}$}
	\email[]{jleddy@txcorp.com}
	\author{B.Dudson$^1$}
	\affiliation{$^1$York Plasma Institute, University of York, Heslington UK}
	\affiliation{$^2$Tech-X Corporation, Boulder CO USA}

	
	
	\date{\today}
	
	\begin{abstract}
		Plasma turbulence is the dominant transport mechanism for heat and particles in magnetized plasmas in linear devices and tokamaks, so the study of turbulence is important in limiting and controlling this transport.  Linear devices provide an axial magnetic field that serves to confine a plasma in cylindrical geometry as it travels along the magnetic field from the source to the strike point.  Due to perpendicular transport, the plasma density and temperature have a roughly Gaussian radial profile with gradients that drive instabilities, such as resistive drift-waves and Kelvin-Helmholtz.  If unstable, these instabilities cause perturbations to grow resulting in saturated turbulence, increasing the cross-field transport of heat and particles.  When the plasma emerges from the source, there is a time, $\tau_{\parallel}$, that describes the lifetime of the plasma based on parallel velocity and length of the device.  As the plasma moves down the device, it also moves azimuthally according to $E\times B$ and diamagnetic velocities.  There is a balance point in these parallel and perpendicular times that sets the stabilisation threshold.  We simulate plasmas with a variety of parallel lengths and magnetic fields to vary the parallel and perpendicular lifetimes, respectively, and find that there is a clear correlation between the saturated RMS density perturbation level and the balance between these lifetimes.  The threshold of marginal stability is seen to exist where $\tau_{\parallel}\approx11\tau_{\perp}$.  This is also associated with the product $\tau_{\parallel}\gamma_*$, where $\gamma_*$ is the drift-wave linear growth rate, indicating that the instability must exist for roughly 100 times the growth time for the instability to enter the non-linear growth phase.  We explore the root of this correlation and the implications for linear device design.
	\end{abstract}
	
	
	\maketitle

\section{Introduction}
Plasmas can be confined in magnetic fields to reduce the transport perpendicular to the field, while leaving the parallel transport unimpeded.  The perpendicular transport levels are then determined by a combination of classical, neoclassical, and turbulent effects - the dominant usually being turbulence.  In the case where turbulence is suppressed, however, the transport is reduced, usually significantly, to neoclassical and classical levels.  Suppression of turbulence is the fundamental feature of H-mode in tokamaks, which is the primary operational regime planned for ITER \cite{Sips2005}.  In a linear device, the geometry of the system means there are no curvature drives for turbulence, so only the pressure gradient provides the free energy - the turbulence is therefore usually seeded by the resistive drift-wave instability \cite{Camargo1995} and the Kelvin-Helmholtz (KH) instability \cite{Popovich2010}.  Drift-waves are perturbations of density and potential that are destabilised by finite parallel resistivity that prevents electrons from perfectly maintaining the Boltzmann relation.  This results in a phase shift between the density and potential perturbations that causes the waves to grow exponentially during the linear regime, until the amplitude is large enough for non-linear effects to dominate, at which point turbulence begins to develop and saturate.  The KH instability is caused by shear flows that begin as laminar, but eddies form to create a mixing layer between the flows eventually destabilising them.  The time during which either these instabilities can develop into turbulence is limited by the parallel transport time from the source to the target in the linear device.  This parallel transport time, as well as the linear growth rates of the resistive drift-wave and KH instabilities, are functions of the parameters of the plasma and the linear device itself (ie. length, magnetic field, density gradients, \emph{etc.}) 

\begin{figure}
	\centering
	\includegraphics[width=\linewidth]{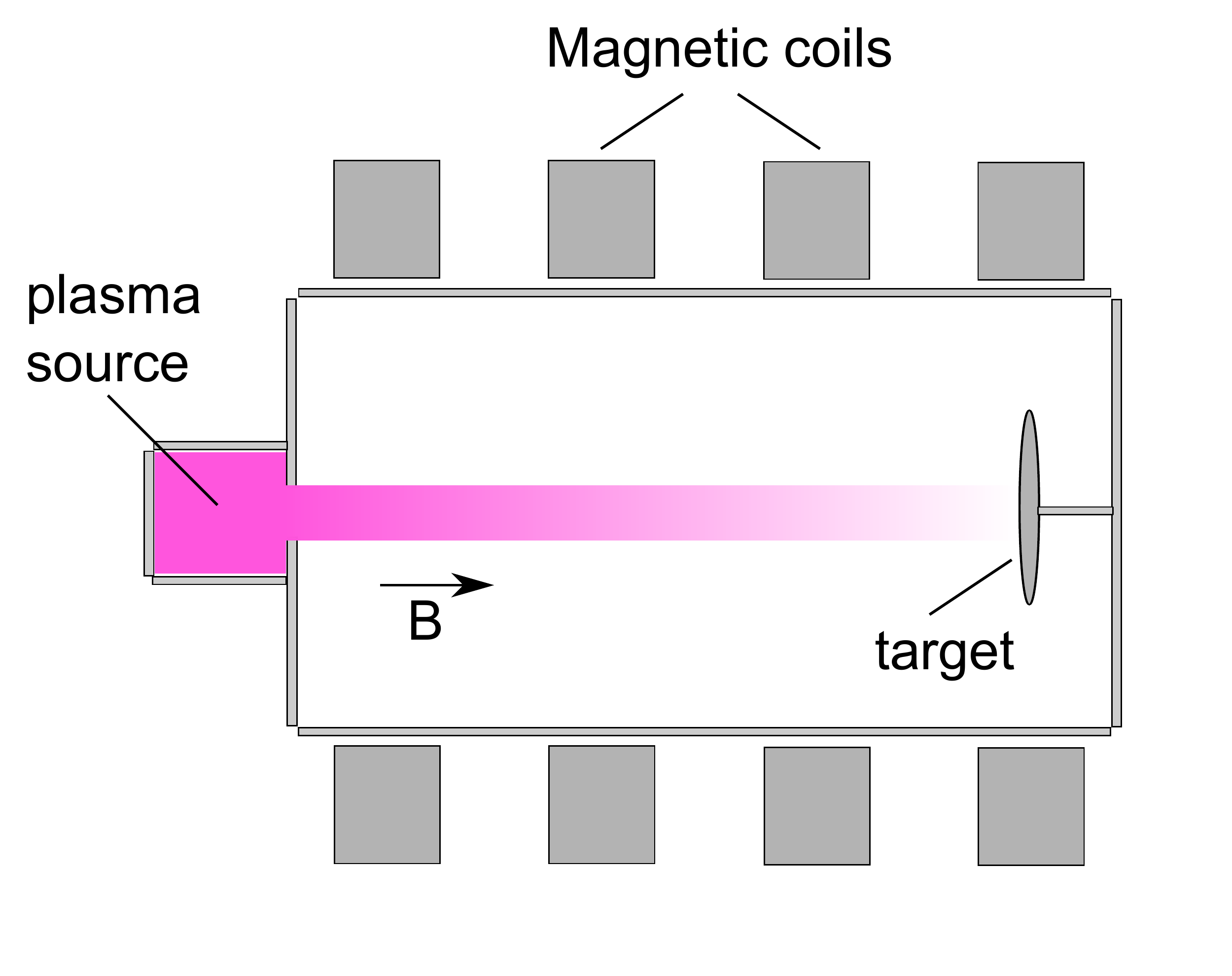}
	\caption{A simple schematic of a linear device.  The magnetic field generated by the coils is roughly constant, pointing to the right.  The plasma source on the left produces a plasma that streams along the field-lines until impacting the target on the right.}
	\label{fig:linear_device}
\end{figure}

Linear devices provide an ideal test-bed for fundamental physics research due to the simple geometry.  They are often used for plasma-material interaction, tokamak edge plasma, and detachment studies \cite{Costin2015,Ohno2017} since they allow direct incidence of the plasma on a target.  The linear device geometry used herein is based on the Magnum-PSI device \cite{DeTemmerman2013} in that the density, temperature, length, and magnetic fields are similar.  The device simulated is defined by a constant axial magnetic field with a plasma source providing plasma density and temperature at one end and a target at the other, as shown in figure \ref{fig:linear_device}.  The plasma source is defined by a Gaussian radial profile in density and temperature with a full-width at half maximum of 8cm.  The peak density and temperature at the plasma source are $10^{19}$m$^{-3}$ and 5eV, respectively.

\section{Plasma simulations}
To simulate the plasma turbulence in linear geometry the Hermes model \cite{Dudson2017} was used, as implemented in BOUT++\cite{Dudson2009}.  This is a 5-field, 2-fluid cold-ion electromagnetic turbulence model that evolves the profiles and fluctuations self-consistently and simultaneously.  The evolution equations are simplified due to the geometry to exclude the curvature terms:
\begin{eqnarray}
\pd{n_e}{t} &=& -\nabla\cdot\left[n_e\left(\mathbf{V}_{E} + \mathbf{b}v_{||e}\right)\right]  \nonumber \\
&& + \nabla\cdot\left(D_\perp\nabla_\perp n_e\right) + S_n
\label{eq:density}
\end{eqnarray}
\begin{eqnarray}
\frac{3}{2}\pd{p_e}{t} &=& -\nabla\cdot\left(\frac{3}{2}p_e\mathbf{V}_{E} + \frac{5}{2}p_e \mathbf{b}v_{||e}\right)  \nonumber \\
&& - p_e\nabla\cdot\mathbf{V}_{E} + v_{||e}\partial_{||}p_e  + \nabla_{||}\left(\kappa_{e||}\partial_{||}T_e\right) \nonumber \\
&& + 0.71\nabla_{||}\left(T_e j_{||}\right) - 0.71 j_{||}\partial_{||} T_e + \frac{\nu}{n_e}j_{||}^2  \\
&&+ \nabla\cdot \left(D_\perp\frac{3}{2} T_e\nabla_\perp n_e\right) + \nabla\cdot\left(\chi_\perp n_e\nabla_\perp T_e\right) + S_p \nonumber
\end{eqnarray}
\begin{eqnarray}
\pd{\omega}{t} &=& -\nabla\cdot\left(\omega\mathbf{V}_{E}\right) + \nabla_{||}j_{||} + \nabla\cdot\left(\mu_\perp\nabla_\perp\omega\right)
\label{eq:vorticity}
\end{eqnarray}
\begin{eqnarray}
\frac{\partial}{\partial t}\left(n_ev_{||i}\right) &=& -\nabla\cdot\left[n_ev_{||i}\left(\mathbf{V}_{E} +  \mathbf{b}v_{||i}\right)\right] - \partial_{||}p_e \nonumber\\
&& + \nabla\cdot \left(D_\perp v_{||i}\nabla_\perp n_e\right)
\label{eq:ion_momentum}
\end{eqnarray}
\begin{eqnarray}
\frac{\partial}{\partial t}\left[\frac{1}{2}\beta_e\psi - \frac{m_e}{m_i}\frac{j_{||}}{n_e}\right] &=& \nu \frac{j_{||}}{n_e} + \partial_{||}\phi - \frac{1}{n_e}\partial_{||} p_e \nonumber \\
&& - 0.71\partial_{||} T_e  \label{eq:ohms_law} \\
&& + \frac{m_e}{m_i}\left(\mathbf{V}_{E} + \mathbf{b}v_{||i}\right)\cdot\nabla\frac{j_{||}}{n_e} \nonumber
\end{eqnarray}
with cross-field E$\times$B and diamagnetic drifts given by
\begin{equation}
\begin{aligned}
\mathbf{v}_E &= \frac{\mathbf{b} \times \nabla \phi}{B} \\ 
\mathbf{v}_D &= \frac{\mathbf{b} \times \nabla p}{enB} \label{eqn:drifts}
\end{aligned}
\end{equation}
respectively.  In this system of equations five fields are evolved where $n_e$ is the plasma density, $p_e$ is the electron pressure, $\omega$ is the vorticity, $v_{\parallel i}$ is parallel ion velocity, $\beta_e$ is the electron beta, $\psi$ is the poloidal flux, and $j_{\parallel}$ is the parallel current.  The plasma simulated herein is chosen to be Deuterium.  The target boundary conditions are set using an insulating sheath with $j_{\parallel}=0$, $\nabla_{\parallel} n=0$, and $v_i\ge c_s$, while the radial boundary conditions are zero gradient.

\begin{figure}
	\centering
	\includegraphics[width=\linewidth]{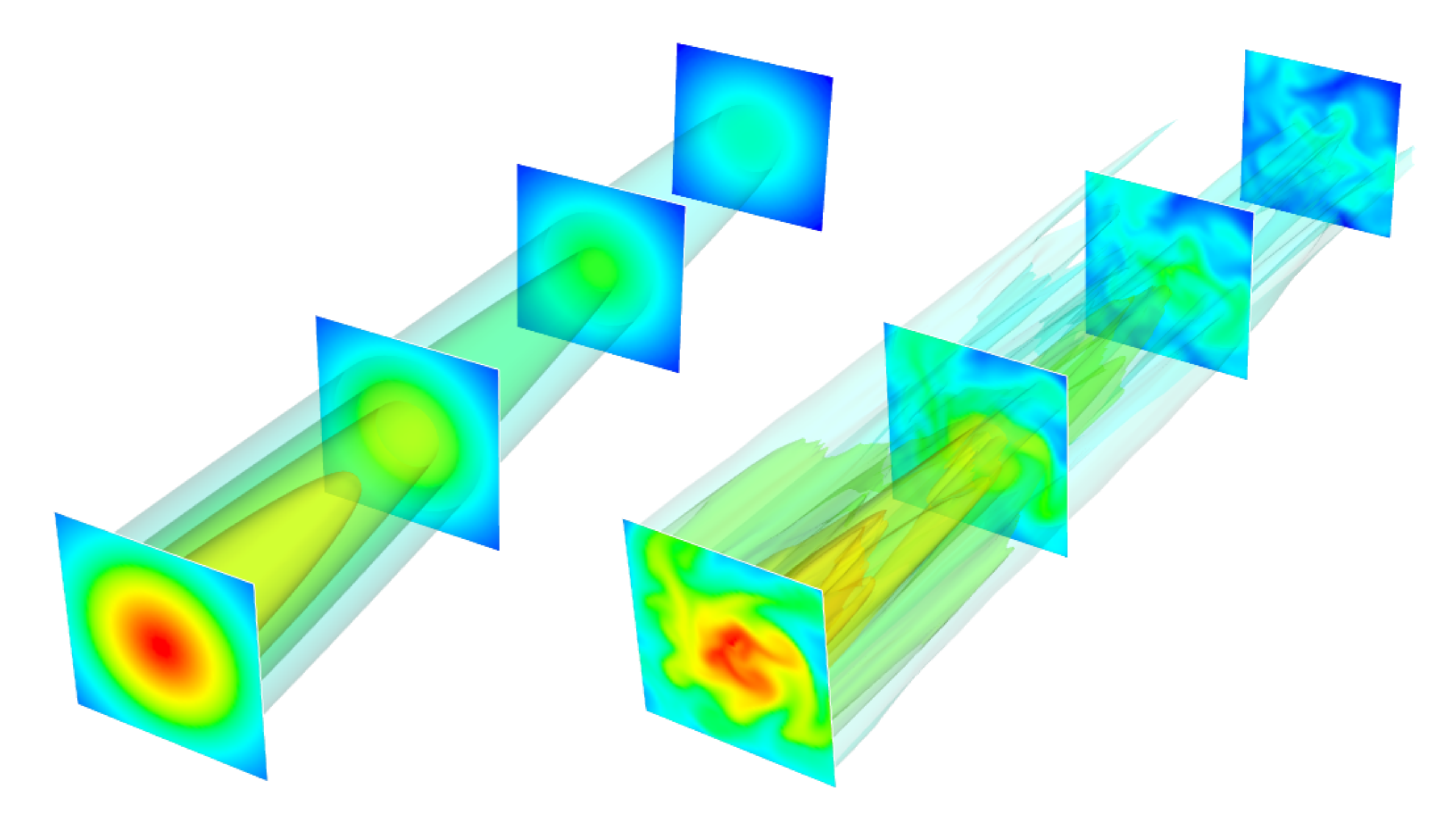}
	\caption{Density contours are shown for two simulations, both of a 1.2m linear device.  The turbulence is suppressed with a magnetic field of $B=0.5$T (left).  The turbulence is unstable and saturates to a fluctuation level of roughly 15\% with the magnetic field of $B=0.1$T (right).}
	\label{fig:comparison_stability}
\end{figure}

\begin{figure}
	\centering
	\includegraphics[width=\linewidth]{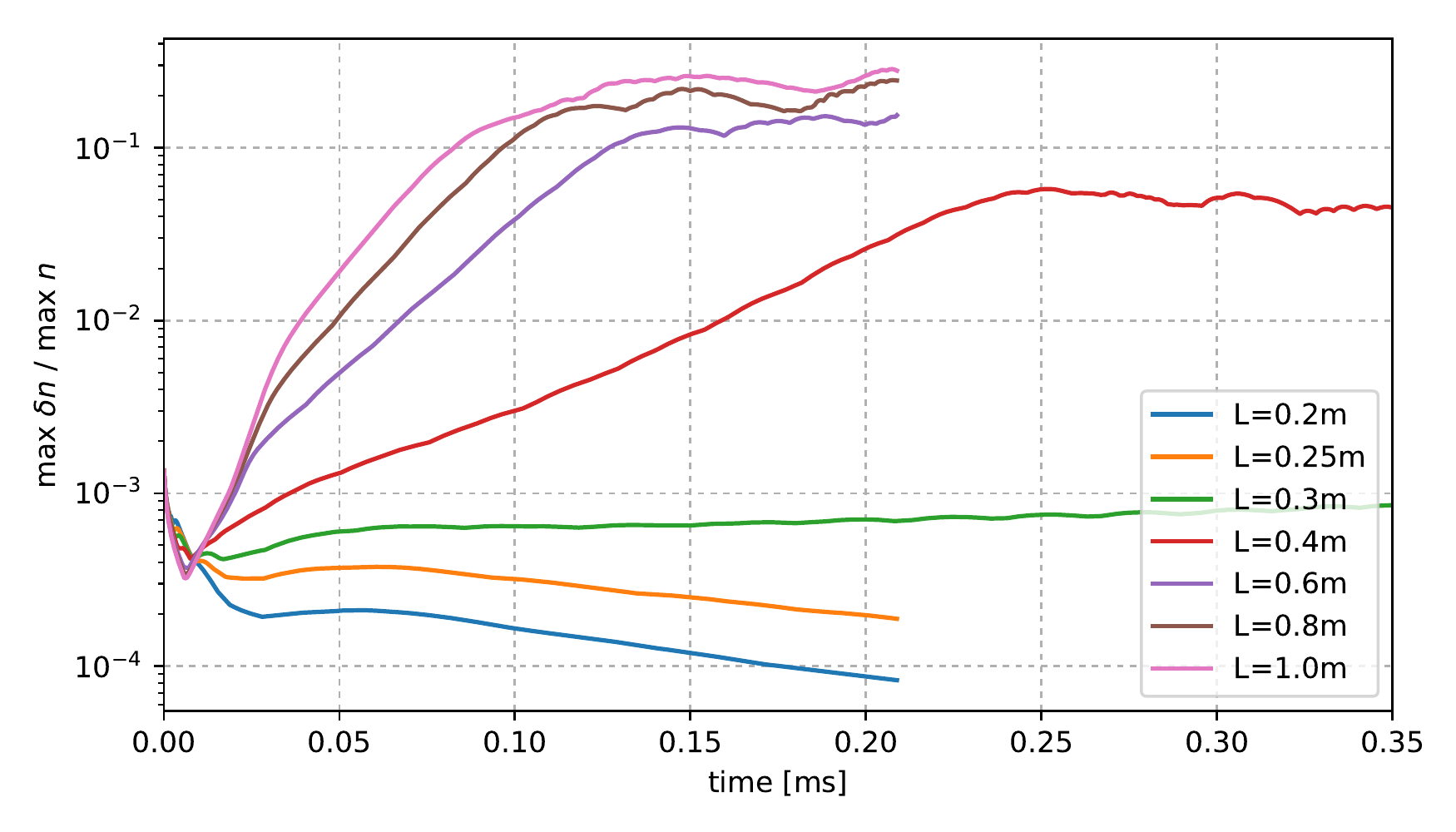}
	\caption{Keeping the magnetic field constant at $B=0.1$T, the length of the device is varied to change the parallel time $L_{\parallel}/v_{\parallel}$.  The maximum density perturbations demonstrate the intrinsic turbulence suppression that occurs as the parallel and perpendicular times satisfy equation \ref{eq:time_balance}.}
	\label{fig:density_perturbations}
\end{figure}

Simulating the plasma in linear geometry with varying magnetic fields led to the discovery of intrinsic turbulence suppression at high field, as shown in figure \ref{fig:comparison_stability}.  Both plots are density contours for simulations of a 1.2m linear device, but with different values of magnetic field.  The time-averaged maximum fractional density perturbation is 15\% for the saturated case at lower magnetic field  $B=0.1$T, but is suppressed to only 0.02\% at the higher magnetic field $B=0.5$T.  Figure \ref{fig:density_perturbations} shows the time evolution of the maximum density perturbations for different length devices.  In all cases the perturbation starts to evolve and grow.  There is a stability threshold somewhere between $L=0.3$m and $L=0.4$m, consistent with the reduced parallel time, $\tau_{\parallel}$, associated with shorter devices.  A relationship clearly exists between the stability of the turbulence and the device length and magnetic field strength. 

\section{Directional balance}
We hypothesize that the stability of the plasma turbulence is dependent on the balance of time scales between parallel and perpendicular (to the magnetic field) motion, reminiscent of critical balance scaling \cite{Barnes2011}.  To examine this condition of marginal stability, a ratio of the time scales is defined:
\begin{equation}
\alpha = \frac{\tau_{\perp}}{\tau_{\parallel}}. \label{eq:time_balance}
\end{equation}
A critical value of this proportionality, $\alpha_c$, represents the threshold of marginal stability.  Assuming there are fundamental velocities ($v_{\perp}$ and $v_{\parallel}$) and distances ($L_{\perp}$ and $L_{\parallel}$) in each direction, this equation is expanded to
\begin{equation}
\alpha = \frac{L_{\perp}v_{\parallel}}{L_{\parallel}v_{\perp}}. \label{eqn:balance}
\end{equation}
The parallel length is simply the length of the linear device.  The perpendicular velocity is composed of two drifts: the $\vec{E}\times\vec{B}$ velocity and the diamagnetic velocity, given by equation \ref{eqn:drifts}.  Because these velocities are both perpendicular to the magnetic field, they can be simplified reducing equation \ref{eqn:balance} to
\begin{equation}
\alpha = \frac{L_{\parallel} \left( en\abs{\nabla_{\perp} \phi} + \abs{\nabla_{\perp} p} \right)}{L_{\perp}enBv_{\parallel}} \label{eqn:Lperp}
\end{equation}
which is an expression for the time-scale ratio, $\alpha$.  For these quantities to be self-consistently calculated, we resort to simulation.  The parallel and perpendicular times are varied by performing parameter scans of two variables: parallel length and magnetic field.  By increasing the length of the device, the parallel time is increased since the average parallel velocity remains relatively constant.  

\section{Stability threshold}
\begin{figure}
	\centering
	\includegraphics[width=\linewidth]{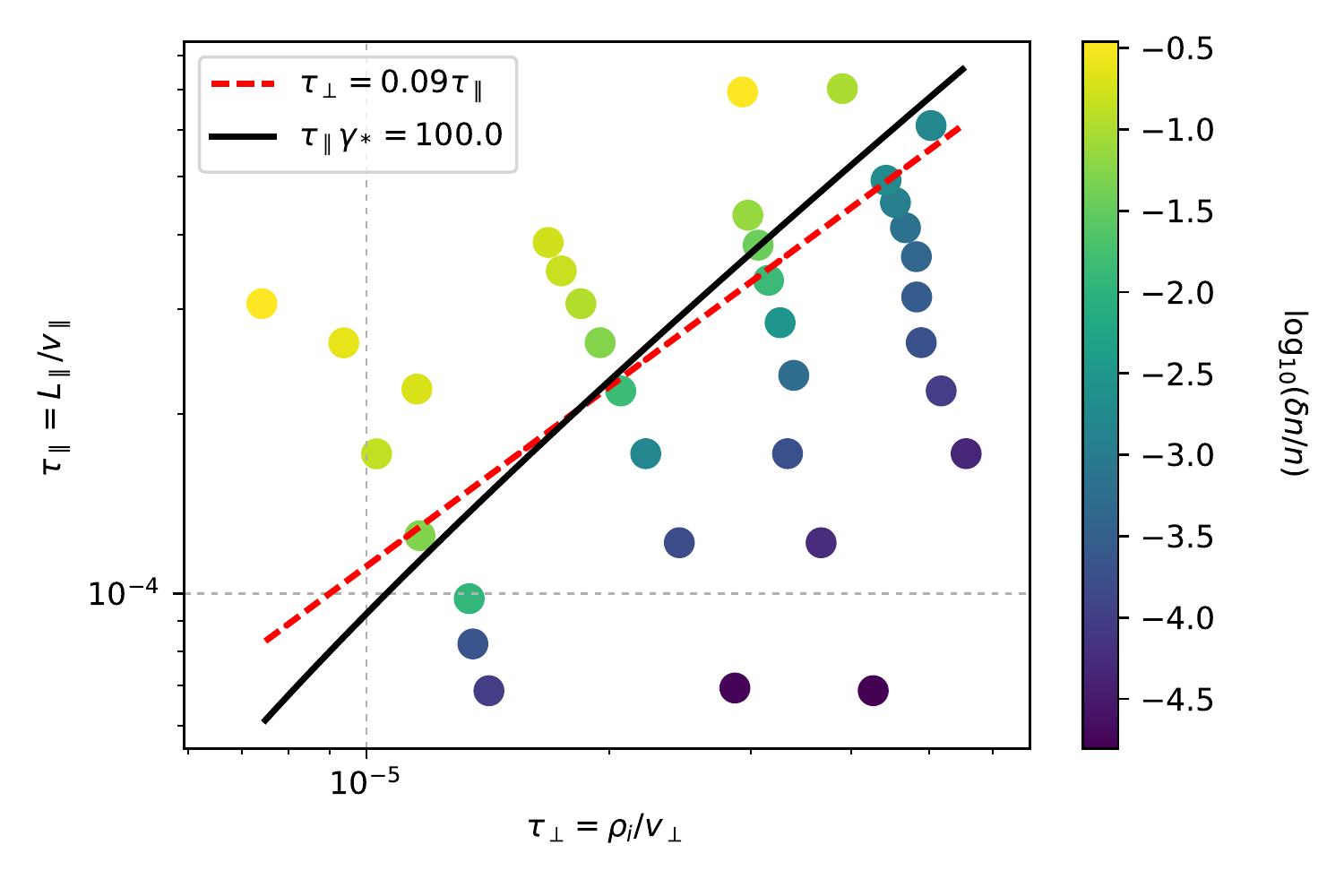}
	\caption{The parameter space $\tau_{\parallel}=L_{\parallel}/v_{\parallel}$ and $\tau_{\perp}=\rho_i/v_{\perp}$ with simulation points showing the RMS log density perturbation of the saturated (or suppressed) turbulence.  The red dashed line indicates where $\tau_{\parallel}\approx10\tau_{\perp}$, which estimates the stability threshold.  This threshold is also well-described by the ratio $\tau_{\parallel}\gamma_*\approx 100$.}
	\label{fig:parameter_space}
\end{figure}

To explore the stability threshold, parameters are varied and simulations run for each case to explore to what degree the turbulence is suppressed.  Figure \ref{fig:parameter_space} shows the parallel vs perpendicular time parameter space, which was explored by varying the magnetic field strength and device length.  Recent simulations of LAPD with external biasing show the development of large azimuthal sheared flows that were found to destabilise the KH instability, which is dominant over the drift-wave instability in seeding the turbulence \cite{Fisher2017}.  In our simulations, the insulating sheath conditions at the target prevent large azimuthal flows from developing, which minimises the impact of the KH instability, so the stability threshold line is calculated using the resistive drift-wave linear growth rate calculated from the slab dispersion relation \cite{Umansky2008,Pecseli2016}: 
\begin{equation}
\left( \omega - \omega_* \right)i\sigma_{\parallel} + \omega^2 = 0  
\end{equation}
where $\omega_*=v_{eth}/L_N$ and
\begin{equation}
\sigma_{\parallel}= \left( \frac{k_{\parallel}}{k_{\perp}} \right)^2 \left[ \frac{\omega_{ci}\omega_{ce}}{0.51 \nu_{ei}} \right]. \label{eqn:spar}
\end{equation}
In these equations $v_{eth}$ is the electron thermal speed, $L_N=n_e\left(\pd{n_e}{r}\right)^{-1}$ is the density scale length, $\nu_{ei}$ is the electron-ion collision frequency, and $\omega_{ci}$ and $\omega_{ce}$ are the ion and electron cyclotron frequencies, respectively. The drift-wave growth rate, $\gamma_* = \text{Im}\; \omega$, is the imaginary part of the complex frequency.  The characteristic perpendicular distance of the plasma is the Larmor radius meaning $L_{\perp} \propto B^{-1}$.  Substituting $k_{\perp} = 2\pi/L_{\perp}$ into equation \ref{eqn:spar} results in a parallel conductivity that is not dependent upon the magnetic field due to the direct dependence of $\omega_{ci}$ and $\omega_{ce}$ on $B$.  This indicates that the drift-wave growth rate is also not a function of magnetic field, therefore constant for all $\tau_{\perp}$.  The black threshold line in figure \ref{fig:parameter_space} is representative of this drift-wave growth time, $\tau_{\parallel}\propto\gamma_*$.  This threshold indicates that the plasma travelling down the linear device in a time $\tau_{\parallel}$ requires at least 100 times the linear drift-wave growth time for turbulence to develop.

\section{Discussion}
\begin{figure}
	\centering
	\includegraphics[width=\linewidth]{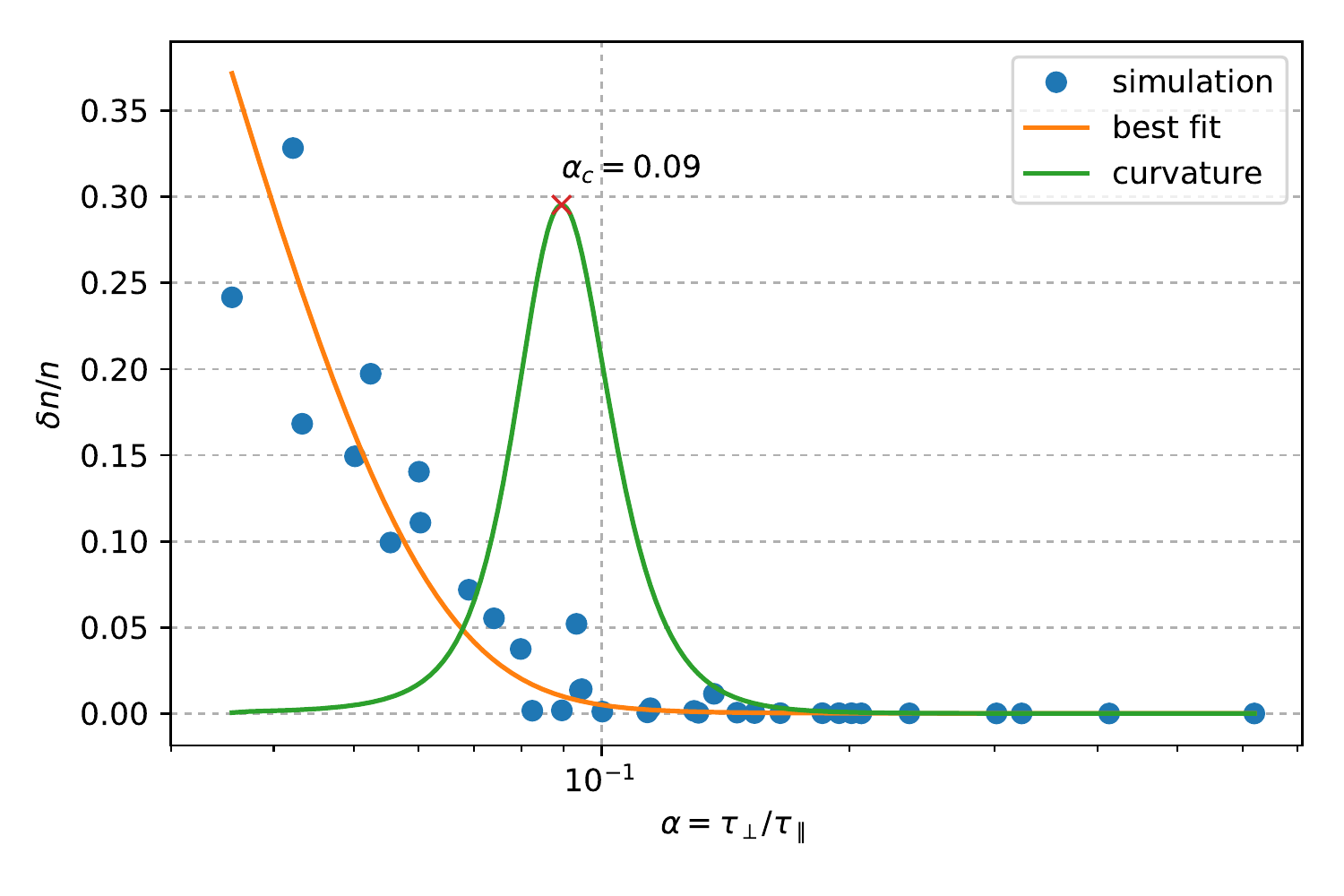}
	\caption{The density perturbation simulation data are plotted against the ratio of perpendicular to parallel times.  This reveals a trend, and a fit allows for a clear threshold value of $\alpha_c = \tau_{\perp}/\tau_{\parallel} = 0.09$ to be found by calculating the curvature, which has a maximum that corresponds to the transition between stable and unstable regimes.}
	\label{fig:dn_vs_alpha}
\end{figure}

To isolate the mechanism for turbulence suppression, a parameter sweep over machine length and magnetic field was conducted producing a surface of $\delta n/n$ in $\tau_{\perp}$ and $\tau_{\parallel}$ space, as shown in figure \ref{fig:parameter_space}.  This figure shows that the density perturbations are directly related the machine length (determining the parallel time $\tau_{\parallel}$) and inversely related to the magnetic field, which determines the perpendicular drift velocity and time, $\tau_{\perp}$.  The line describing $\alpha=\tau_{\perp}/\tau_{\parallel}=0.09$ can be seen more clearly in figure \ref{fig:dn_vs_alpha}, where the density perturbation amplitude is plotted versus the stability threshold parameter, $\alpha$.  A clear trend is seen - turbulence is destabilised at low $\alpha$ and suppressed at high $\alpha$.  The transition between these two regimes is relatively sharp, a feature highlighted by figure \ref{fig:dn_vs_alpha}.  A critical value of $\alpha_c$ can be found for the stability threshold by looking at the curvature of the best fit line.  The maximum curvature represents the transition between the stable and unstable regimes.  The threshold value $\alpha_c=0.09$ also closely corresponds to the 1\% fluctuation level.

In the simulations it appears that the mechanism for suppressing the turbulence is the shearing of perturbations by parallel and perpendicular flows due to the local and global potential and pressure gradients.  Due to the gaussian nature of radial temperature profile, the sound speed at the sheath is low at high radius.  Since the parallel flows is at least Mach 1 at the sheath, an intrinsic radial shear develops in the parallel flows shearing the density perturbations.

The implications for linear device design are clear: magnetic field and linear device length can be used as tunable parameters that can either stabilise or destabilise turbulence depending on device purpose.  Linear devices such as LAPD, which are very long with relatively low magnetic field, observe saturated turbulence as expected \cite{Pueschel2017}.  Experimental evidence has shown that drift-waves can be suppressed in the presence of an external potential bias imposing a radial electric field on the plasma \cite{Dubois2012}; however, a comprehensive scan of magnetic field and device length has not to our knowledge been conducted.  This work therefore serves as a theoretical prediction of the plasma behaviour in regards to resistive drift-wave turbulence stability in linear devices.

\subsubsection*{Acknowledgements}
Some of the simulations were run using the ARCHER computing service through the Plasma HEC Consortium EPSRC grant number EP/L000237/1.   This work has received funding from the RCUK Energy Programme, grant number EP/M001423/1.  It has been carried out within the framework of the EUROfusion Consortium and has received funding from the Euratom research and training programme 2014-2018 under grant agreement No 633053.  The views and opinions expressed herein do not necessarily reflect those of the European Commission.

\bibliographystyle{unsrt} 
\bibliography{master}

\end{document}